\documentclass[english,12pt,nofootinbib,preprintnumbers,showpacs,eqsecnum,prd]{revtex4}

\usepackage{graphicx}
\usepackage{amssymb}
\usepackage{babel}
\usepackage{epsfig}
\usepackage{latexsym,amsmath,tabularx,amssymb,bm}
\usepackage{graphicx}

\def\q{{\boldsymbol q}}

\newcommand \beq{\begin{eqnarray}}
\newcommand \eeq{\end{eqnarray}}
\def\simle{\mathrel{\rlap{\raise 0.511ex \hbox{$<$}}{\lower 0.511ex 
\hbox{$\sim$}}}}
\def\simge{\mathrel{ \rlap{\raise 0.511ex 
\hbox{$>$}}{\lower 0.511ex \hbox{$\sim$}}}}
\newcommand{\del}{\partial}

\begin{document}


\title{Calculation of the pressure of a hot scalar theory within the Non-Perturbative Renormalization Group }

\author{Jean-Paul Blaizot}
\affiliation{IPhT, CEA Saclay 91191 Gif-sur-Yvette cedex, France}

\author{Andreas Ipp}
\affiliation{Institut f\"ur Theoretische Physik,
Technische Universit\"at Wien\\
Wiedner Hauptstra{\ss}e 8-10, 1040 Vienna, Austria}

\author{Nicol\'as Wschebor}
\affiliation{Instituto de F\'\i sica, Facultad de Ingenier\'\i a,\\
J.H.y Reissig 565, 11000 Montevideo, Uruguay}

\date{July 6, 2010}

\begin{abstract}
We apply to the calculation of the pressure of a hot scalar field theory  a method that has been recently developed to solve  the Non-Perturbative  Renormalization Group. This method yields an accurate determination of the momentum dependence of $n$-point functions over the entire momentum range, from the low momentum, possibly critical, region up to the perturbative, high momentum region. It has therefore the potential to account well for the contributions of modes of all wavelengths to the thermodynamical functions, as well as for the effects of the mixing of quasiparticles with multi-particle states.  We compare the thermodynamical functions obtained with this method to those of the so-called Local Potential Approximation, and we find extremely small corrections.   This result points to the robustness of the quasiparticle picture in this system. It also demonstrates the stability of the overall approximation scheme, and this up to  the largest values of the coupling constant that can be used in a scalar theory in 3+1 dimensions. This is in sharp contrast to perturbation theory which shows no sign of convergence, up to the highest orders that have been recently calculated.
\end{abstract}

\maketitle


\section{Introduction}

There has been much effort devoted in the recent years to the development of finite temperature field theory, in particular in the context of Quantum Chromodynamics (QCD) whose thermal properties are central to the understanding of heavy ion collisions at high energy. Equilibrium properties of hot QCD are calculable from lattice gauge theory (for a review, see e.g.  \cite{Karsch:2003jg}), but there is a need to develop semi-analytical tools to understand the results of such calculations, with the hope that such tools may also allow one to approach non-equilibrium situations. 

Weak coupling expansions are among such tools. In the case of QCD, the use of perturbation theory is  motivated by the asymptotic freedom that leads to a small effective coupling at high temperature.  However, 
strict perturbation theory does not work at finite temperature: it exhibits indeed very poor convergence properties, even in a range of values of the coupling where good results are obtained at $T=0$. This difference of behavior of perturbation theory at zero and finite temperature can be  understood from the  fact that, at finite temperature, the expansion parameter involves both the coupling and the magnitude of thermal fluctuations  (for a recent review, see
\cite{Blaizot:2003tw}; see also \cite{Blaizot:2009iy}).  In that respect, the problem is not specific to QCD: Similar poor convergence behavior appears also
in the simpler scalar field theory \cite{Parwani:1995zz}, and has also
been observed in the case of large-$N$ $\phi^{4}$ theory
\cite{Drummond:1997cw}. Similar observations can be made also in the case of QED \cite{Andersen:2009tw}.

In this paper we restrict ourselves to the case of the simple theory of a scalar field $\varphi$, with a quartic self-coupling $g^2\varphi^4$, for which high order calculations were recently completed. Thus the pressure is  known  to order $g^8 \ln g$ \cite{Andersen:2009ct}, and the screening mass to order $g^6 \ln g$ \cite{privateAndersen}.  Such calculations were made by exploiting effective theory techniques, in particular dimensional reduction, that rely on a separation of scales of various degrees of freedom in the hot scalar plasma: 
hard modes with momenta $k\sim T$ that contribute dominantly to the pressure and are weakly coupled, and soft modes with momenta $k\sim gT$ that are more strongly coupled. In QCD, another scale concerns the ultrasoft modes, with momenta $k\sim g^2T$ that remain strongly coupled for any value of the coupling constant. This scale, relevant only in the case where massless modes exist at finite temperature,  does not play any role in the scalar field theory.   
 The separation of scales that  allows the organization of the calculation using effective field theory disappears when  the coupling is not too small: then the various degrees of freedom mix and the situation requires a different type of analysis. The purpose of this paper is to
apply to this problem the Non-Perturbative Renormalization
Group (NPRG)  (for reviews of this method see e.g. 
\cite{Bagnuls:2000ae,Berges:2000ew,Delamotte:2007pf}). 

To do so, we shall rely on an elaborate approximation scheme that has been developed recently in order to obtain a good determination of the momentum dependence of the $n$-point functions \cite{Blaizot:2005xy}. This new approximation has been  tested on the $O(N)$ models, for which it provides excellent critical exponents, and more generally, an excellent description of the momentum dependence of the 2-point function, from the low momenta of the critical region, all the way up to the large momenta of the perturbative regime \cite{Benitez:2009xg}. One may then expect this method to capture accurately the contributions to the thermodynamical functions  of thermal fluctuations  from various momentum ranges, and hence handle properly the mixing between degrees of freedom that takes place as the coupling grows. Since it involves also non trivial momentum dependent self-energies, the method  also encompasses effects related to the damping of quasiparticles, or their coupling to complex multi-particle configurations. 

The present  paper may be viewed, in its spirit and goals, as a follow up of the analysis presented in Ref.~\cite{Blaizot:2006rj}. However it departs from it  in two ways. The first difference is of a technical nature: the new approximation scheme is better justified  when one uses an Euclidean symmetric four dimensional regulator  that cuts off the contribution of high Matsubara frequencies. In contrast, in Ref.~\cite{Blaizot:2006rj}, we used a three dimensional regulator, and performed analytically the (untruncated) sums over the Matsubara frequencies. The second difference is that we use a new, much more accurate,  approximation scheme to solve the NPRG equations, as mentioned above: The Local Potential Approximation (LPA) used in Ref.~\cite{Blaizot:2006rj} can be viewed as  the zeroth order in this new approximation scheme. In physical terms, the LPA corresponds to an approximation where the degrees of freedom of the hot scalar plasma are massive quasiparticles. The new scheme goes beyond that simple picture. 
As it turns out, the results obtained are not too different from those of the LPA. This stability of the results against improvements in the approximation suggests that the scheme that we are using to solve the  NPRG equations may give already, at the level at which it is implemented here, an accurate representation of the exact pressure, and this over a wide range of coupling constants. It also indicates that for such a system the quasiparticle picture is presumably robust. 

  The outline of this paper is as follows. In Sec.~\ref{sec:bmw}, we
present a brief introduction to the NPRG and the  approximation scheme of Ref.~\cite{Blaizot:2005xy}, indicating specific features of finite
temperature calculations.  In Sec.~\ref{sec:numerical} we integrate the flow
equations numerically and discuss the  results obtained.
  The last section summarizes the conclusions. In the Appendices we give details about the numerical integration needed to calculate the flow of the pressure, and we discuss specific features of the exponential regulator used in our calculations at finite temperature. 


\section{The NPRG in the  BMW approximation scheme}
\label{sec:bmw}

We consider  a scalar field theory with
the classical (Euclidean) action \beq\label{eactON} S = \int_0^{1/T} {\rm d}\tau\int{\rm
d}^{3}x\,\left\lbrace{ \frac{1}{2}}   \left(\del
\varphi(x)\right)^2  + \frac{m^2}{2} \, \varphi^2(x) + \frac{u}{4!}
\,\varphi^4(x) \right\rbrace \,, \eeq
where $T$ is the temperature. Our goal is to calculate the thermodynamical pressure for this scalar theory, using the NPRG. We follow here   Ref.~\cite{Berges:2000ew}, and add 
   to the original  
 action $S$ a regulator   
\beq\Delta S_\kappa[\varphi]= \frac{1}{2}\int_q\: R_\kappa(q)\varphi(q)\varphi(-q),\eeq 
where the parameter $\kappa$ runs continuously from the  microscopic scale $\Lambda$ down  to 0. 
We use the  notation
 \beq
\int_q\equiv T\sum_{\omega_n}\int  \frac{d^3q}{(2\pi)^3},
\eeq
 with $q=(\omega_n,\q)$,  $q^2=\omega_n^2+\q^2$, and $\omega_{n}=2\pi nT$ are  the Matsubara frequencies.  The role of $\Delta S_\kappa$
is to suppress the fluctuations with momenta $|q|\lesssim \kappa$,
while leaving unaffacted those with $|q| \gtrsim \kappa$ ($|q|=\sqrt{q^2}$). This is achieved with a cut-off function  $R_\kappa(q)$ that has the following properties: 
$R_\kappa(q)\rightarrow \kappa^2$ when $|q| \ll \kappa$, and 
$R_\kappa(q)\rightarrow 0$ when $|q|\gg \kappa$. The precise form of the function $R_\kappa(q)$ used in our calculation will be discussed later.

The effective action 
$\Gamma_\kappa[\phi ]$ associated to the action $S+\Delta S_\kappa$ obeys the exact flow equation \cite{Wetterich:1992yh} (with $\del_t\equiv \kappa\del_\kappa$):
 \beq \label{NPRGeq}
\partial_t \Gamma_\kappa[\phi]=\frac{1}{2}\int_q
\,\partial_t R_\kappa(q)\,
G_\kappa[q,-q;\phi],
\eeq
where $G_\kappa[q,-q;\phi]$ is the full propagator in the presence of the background field $\phi$:
\beq
G_\kappa^{-1}[q,-q;\phi]=\Gamma_\kappa^{(2)}[q,-q;\phi]+R_\kappa(q),
\eeq
with $\Gamma^{(2)}_\kappa[q,-q;\phi]$  the second functional derivative of 
$\Gamma_\kappa[\phi]$ w.r.t. $\phi$. 
The initial condition of the flow is specified at 
the microscopic scale $\kappa=\Lambda$: at this point, we assume that the fluctuations are completely frozen by  
$\Delta S_\kappa$, so that $\Gamma_{\kappa=\Lambda}[\phi]\approx S[\phi]$. 
The effective action $\Gamma[\phi]$ of the scalar field theory  is obtained as the solution of  Eq.~(\ref{NPRGeq})
for $\kappa\to 0$, at which point $R_\kappa(q)$ vanishes.

When $\phi$ is constant, the functional $\Gamma_\kappa[\phi]$  reduces, to within a volume factor, to the effective potential $V_\kappa(\phi)$.
The  flow equation for  $V_\kappa$ follows from that of the effective action $\Gamma_\kappa$, Eq.~(\ref{NPRGeq}),  when restricted to a constant $\phi$. It reads   \beq\label{eqforV} \del_t
V_\kappa(\rho)= \frac{1}{2}\int_q \del_t R_\kappa(q)\, G_\kappa(q,\rho), \eeq
where 
\begin{equation}\label{G-gamma2}
G^{-1}_{\kappa} (q,\rho) \equiv \Gamma^{(2)}_{\kappa} (q,\rho) +
R_\kappa(q),\qquad 
\rho\equiv \frac{\phi^2}{2}.
\end{equation}
We  used here the simplified notation 
 $\Gamma^{(2)}_{\kappa} (q,\rho)$ in place of $\Gamma^{(2)}_{\kappa}
(q,-q,\rho)$ for the $2-$point function in a constant background field, and similarly for $G(q,\rho)$. Also, we have set $\rho\equiv \phi^2/2$, a notation to be used throughout (when $\phi$ is constant). The pressure $P$ is related to the effective potential by
\beq\label{pressuredef}
P_\kappa (T)=-\left[  V_\kappa(T,\rho=0)-V_\kappa(T=0,\rho=0) \right].
\eeq

The equation for the effective potential may be viewed as the equation for the ``zero-point'' function in a constant background field. By taking two derivatives with respect to $\phi$ and letting $\phi$ be constant, one obtains the 
equation for the 2-point function in a constant background field:
\begin{eqnarray}
\label{gamma2champnonnul}
\partial_t\Gamma_{\kappa}^{(2)}(p)&=&\int_q
\,\partial_t R_\kappa(q)\,G_{\kappa}^2(q)\nonumber\\
&\times&\left\{\Gamma_{\kappa}^{(3)}(p,q,-p-q) G_{\kappa}(q+p)\Gamma_{\kappa}^{(3)}(-p,p+q,-q)
-\frac{1}{2}\Gamma_{\kappa}^{(4)}
(p,-p,q,-q)\right\} .\nonumber \\\end{eqnarray}
In this equation, all the $n$-point functions depend on the constant background field $\phi$.
This has not been indicated explicitly in order to alleviate the notation.

The flow equations (\ref{eqforV}) and (\ref{gamma2champnonnul}) are the first equations of an infinite tower of coupled equations for the $n$-point functions, whose solution requires some truncation. We shall use the truncation scheme proposed recently by Blaizot, M\'endez-Galain and Wschebor (BMW) \cite{Blaizot:2005xy},  implemented here at its lowest non-trivial order: in this case we need only consider the equation for the effective potential and that for the 2-point function, that is, Eqs.~(\ref{eqforV}) and (\ref{gamma2champnonnul}), on which we shall perform approximations that are described next. For a more complete discussion we refer to Ref.~\cite{Benitez:2010aa}. 

The lowest level of the approximation scheme corresponds to a widely used approximation, usually referred to as the Local Potential Approximation (LPA). It consists in assuming that for all values of $\kappa$ the effective action takes the form \cite{Berges:2000ew}
\beq\label{GammaLPA}
\Gamma_\kappa[\phi]=\int_0^\beta \int d^3x\left\{ \frac{1}{2} \left( \del\phi\right)^2 +V_\kappa(\phi)\right\},
\eeq
which is tantamount to assume that the 2-point function is of the form
\beq\label{GLPA}
\Gamma^{(2)}_\kappa (q,\rho)=q^2+m_\kappa^2(\rho),\qquad m_\kappa^2(\rho)\equiv\partial_\phi^2
V_\kappa.
\eeq
With this Ansatz, Eq.~(\ref{eqforV})  for $V_\kappa$  becomes a closed equation that we write as follows
 \beq\label{VeffI1}
\partial_t
V_\kappa(\rho)=\frac{1}{2}I_1,
\eeq
where $I_1$ is one of the following integrals 
\beq\label{defIJ}
J_n(p )\equiv
\int_q \;\partial_t R_\kappa(q)\;
G_\kappa(p+q )G^{n-1}_\kappa(q ), \qquad I_n \equiv J_n(p=0). \eeq
Note that Eq.~(\ref{VeffI1}) is formally an exact equation, and it would yield the exact effective potential  if $I_1$ were calculated with the exact propagator. In the LPA, the equation keeps the same form, but the propagator is given by Eq.~(\ref{GLPA}), where $m^2_\kappa(\rho)$ is itself determined by the potential (thereby making Eq.~(\ref{VeffI1}) a closed, self-consistent equation). 

The next order of the approximation scheme, to which we refer to  as the leading order of the BMW method  \cite{Blaizot:2005xy}, or here  simply as BMW for briefness,  consists in neglecting  the loop momentum $q$ in the 3 and 4-point functions in the right hand side of Eq.~(\ref{gamma2champnonnul}).  Once this approximation is made,  the
corresponding $n$-point functions can be obtained as the derivatives
of  the 2-point function with respect to the constant
background field: \beq\label{derivs}
\Gamma_{\kappa}^{(3)}(p,-p,0,\phi)=\partial_\phi\Gamma_{\kappa}^{(2)} (p,\phi) , \hskip 1 cm
\Gamma_{\kappa}^{(4)}(p,-p,0,0,\phi)=\partial_\phi^2
\Gamma_{\kappa}^{(2)} (p,\phi). \eeq 
The equation for the 2-point function becomes then a closed equation
 \beq
 \label{2pointcloseda}
\partial_t\Gamma_\kappa^{(2)}(p,\rho)=
J_3(p) \; \left( \partial_\phi
\Gamma_\kappa^{(2)}(p,\rho)\right)^2
  -\frac{1}{2} I_2 \; \partial_\phi^2
\Gamma_\kappa^{(2)}(p,\rho).
\eeq
There is however a subtlety:  this equation for the 2-point function is coupled to that of the effective potential,  because $\Gamma_\kappa^{(2)}(p=0,\rho)=\partial_\phi^2 \, V_\kappa(\phi)$. In order to properly implement this coupling, we 
treat separately the zero momentum ($p=0$) and the non-zero momentum
($p\ne 0$) sectors, and  
define
\beq\label{defsigmabis}
\Gamma^{(2)}_\kappa (p,\rho)\equiv  p^2+ \Delta_\kappa (p,\rho)+m_\kappa^2(\rho),
\eeq
where $m_\kappa^2(\rho)=\partial_\phi^2 V_\kappa(\rho)$ is obtained by solving the equation for the effective potential. The equation for $\Delta_\kappa(p,\rho)$ can be easily deduced from that for $\Gamma^{(2)}$, i.e., from Eq.~(\ref{2pointcloseda}) by subtracting the corresponding equation that holds for $p=0$. It reads
\beq\label{2pointclosedab0}
 \partial_t\Delta_\kappa(p,\rho)&=&
2\rho  J_3(p,\kappa,\rho) \; \left[u_\kappa(\rho)+\Delta_\kappa^\prime(p,\rho)\right]^2-2\rho I_3(\kappa,\rho)\; u_\kappa^2(\rho)  \nonumber\\
  &-&\frac{1}{2} I_2(\kappa,\rho) \; \left[
\Delta_\kappa^\prime(p,\rho)+2\rho
\Delta_\kappa^{\prime\prime}(p,\rho)\right],
\eeq
where the symbol $^\prime$ denotes the derivative with respect to $\rho$, and we have set $u_\kappa(\rho)\equiv\del m_\kappa^2(\rho)/\del \rho$.

This equation (\ref{2pointclosedab0}), together with Eq.~(\ref{eqforV}) for the effective potential 
and that for the propagator
\beq
G_\kappa^{-1}(q,\rho)=q^2+\Delta_\kappa(q,\rho)+m_\kappa^2(\rho)+R_\kappa(q),\eeq
constitute a closed system of equations for $\Delta_{\kappa}(p,\rho)$ and $V_\kappa(\phi)$. This  can be solved with the initial condition $\Gamma_{\Lambda}^{(2)}(p;\rho)=p^2+m^2+u \rho$, where  $m^2$ and $u$ are essentially (to within small ultraviolet cut-off corrections) the parameters of the action (\ref{eactON}).

 We now specify the regulator $R_\kappa(q)$ that we have used in our calculation.  We take it of the generic (Euclidean symmetric) form 
\beq \label{regulator}
R_\kappa(q)=Z_\kappa \kappa^2 r(\tilde q),\qquad \tilde q\equiv \frac{q}{\kappa},
\eeq
where $Z_\kappa$ is a function of $\kappa$ only, to be specified shortly, and the function $r(\tilde q)$ is a smooth function of its argument.  In most of our calculations, we have used an exponential regulator of the form
\beq \label{regulator1}
r(\tilde q)=\frac{\alpha \tilde q^2}{{\rm e}^{\tilde q^2}-1},\qquad \tilde q=\frac{q}{\kappa},
\eeq
where $\alpha$ is a free parameter.
We have also considered an alternative regulator, given by
\begin{equation} 
r(\tilde q)=\alpha e^{-\beta \tilde q^2 - \gamma \tilde q^4},\label{regulator2}
\end{equation}
with parameters $\beta=1/2$ and $\gamma=1/24$ chosen so that  the Taylor expansion around $\tilde q = 0$ agrees with the  regulator (\ref{regulator1}) through order $O(\tilde q^4)$, and leave only the prefactor $\alpha$ as a free parameter \footnote{We are grateful to B. Delamotte and H. Chat\'e for suggesting the usefulness of this regulator}. The regulators (\ref{regulator1}) and (\ref{regulator2}) are suited for the BMW  approximation whose justification relies on both momenta and frequencies being cut-off. For comparison, we have also solved the LPA with these exponential regulators. We shall also compare to  the LPA results of Ref.~\cite{Blaizot:2006rj} where the Litim regulator $r(\tilde q)=(1-\tilde q^2) \theta(1-\tilde q^2)$ \cite{Litim:2000ci} was used in integrals over three momenta, the Matsubara frequencies being integrated over analytically.

In the absence of any approximation, the physical quantities such as the pressure should be strictly independent of the 
cut-off function, and in particular of the value of the parameter $\alpha$ that we have introduced in the regulators (\ref{regulator1}) and (\ref{regulator2}). In practice, we find a (weak) residual dependence on the parameter $\alpha$, and 
 a study of this spurious dependence  provides an indication of the quality of the 
 approximation. 
 It should be emphasized however that this dependence on $\alpha$ is rather small, as we shall see, and we have made no effort to perform a systematic study of the dependence of our results on its value. Nor did we explore the effects of enlarging the space of possible variations by allowing for instance the parameters $\beta$ and $\gamma$   in (\ref{regulator2}) to take arbitrary values.

The factor $Z_\kappa$ in Eq.~(\ref{regulator})   reflects the finite change in normalization of the field between the scale $\Lambda$ and the scale $\kappa$. It is defined by
\begin{equation}\label{Zkapparho}
Z_\kappa=\left.\frac{\partial }{\partial p^2}\Gamma^{(2)}_{\kappa}(p,\rho)\right|_{p=0, \rho=\rho_0},
\end{equation}
where $\rho_0$ is a priori arbitrary but chosen here to correspond to the minimum of the effective potential. This  factor $Z_\kappa$ enters also the definition of the dimensionless variables that are used in the numerical solution. Thus, for instance we define
\beq\label{dimensionless}
q\equiv \kappa\tilde q,\qquad \rho\equiv K \kappa^{2} Z_\kappa^{-1}\tilde\rho,\qquad m_\kappa^2\equiv Z_\kappa \kappa^2\tilde m_\kappa^2,\qquad u_\kappa\equiv  Z_\kappa^2 K^{-1}\tilde u_\kappa, 
\eeq
with $m_\kappa\equiv m_k(\rho=0)$,  $u_\kappa\equiv u_\kappa(\rho)$, and $K\equiv 1/(32\pi^2)$. In the next section, when we discuss the results obtained by solving numerically Eqs.~(\ref{2pointclosedab0}) and (\ref{eqforV}), we shall set $g_\kappa^2\equiv u_\kappa/24$.



\section{Results}
\label{sec:numerical}

To calculate temperature dependent physical quantities, we need to evaluate the flow of these quantities at both zero temperature and at finite temperature. We follow here the strategy exposed in Refs.~\cite{Tetradis:1992xd,Blaizot:2006rj}. 
Conceptually, one starts
with given physical parameters at zero temperature at the scale $\kappa=0$.
One then removes quantum
fluctuations step by step by integrating the flow
equations upwards from $\kappa=0$ to $\Lambda$ in order to arrive at ``bare quantities'' at a
chosen scale $\Lambda$. If $\Lambda$ is chosen big enough, only
renormalizable  local operators survive in the effective action, and the
system can then be described by a simple set of bare parameters $g_\Lambda$, $m_\Lambda$, and $Z_\Lambda$. Starting
from these bare parameters one then follows the flow downwards  from
$\kappa=\Lambda$ to 0, but this time with the temperature $T$ turned on.
The physical quantities are then obtained at
$\kappa=0$.

In practice, for reasons of numerical stability, the flow is never integrated upward,
but always downward, for both zero and finite temperature.
For a given bare coupling $g_\Lambda$ (and $Z_\Lambda=1$), one adjusts the bare mass $m_\Lambda$
at the scale $\kappa=\Lambda$ by bisection, so as to arrive
at a vanishing 
mass $m_0=0$ at
the end ($\kappa=0$) of the zero temperature flow,  to the desired accuracy.
Keeping the same bare parameters, one  then turns on the temperature and runs 
the flow again. Note that we consider specifically here  a massless theory in order to be able to compare our results with the known results of  high order perturbation theory. But the same analysis could be done for any value of $m_0$.

Since the zero-temperature coupling constant vanishes at $\kappa=0$ for any $g_\Lambda$ (the so-called triviality of $\varphi^4$ theory in $d=4$), one 
adjusts the coupling constant at a finite scale $\kappa > 0$.
In accordance with what is commonly done in perturbation theory, we choose to do that at 
$\kappa=2\pi T$ on the $T=0$ flow. This procedure induces a specific
scheme dependence, attached to the choice of the regulator, which should be taken into consideration 
when comparing with results that are obtained in another scheme,
for example the minimal subtraction scheme in perturbation theory. 

In order to keep the zero-temperature flow under control, 
we introduce
dimensionless variables \cite{Berges:2000ew}.
Although these are not optimal for the thermal flow, which freezes in dimensionful
variables (but diverges
in dimensionless variables) as $\kappa \rightarrow 0$,
it is still advantageous to use the same scaling of variables as for the
zero-temperature flow in order to achieve high-precision cancellations between zero
and finite temperature flows for larger $\kappa \gg T$.

\begin{figure}
\hfill{}\includegraphics[scale=0.65]{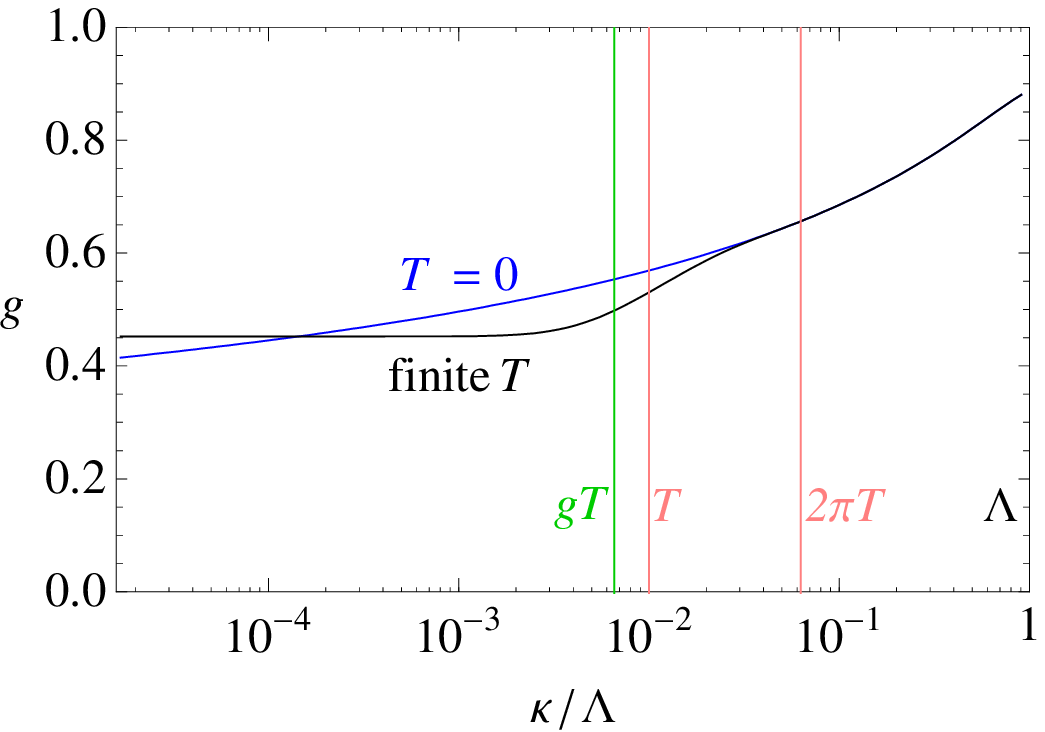}\hfill{}\includegraphics[scale=0.65]{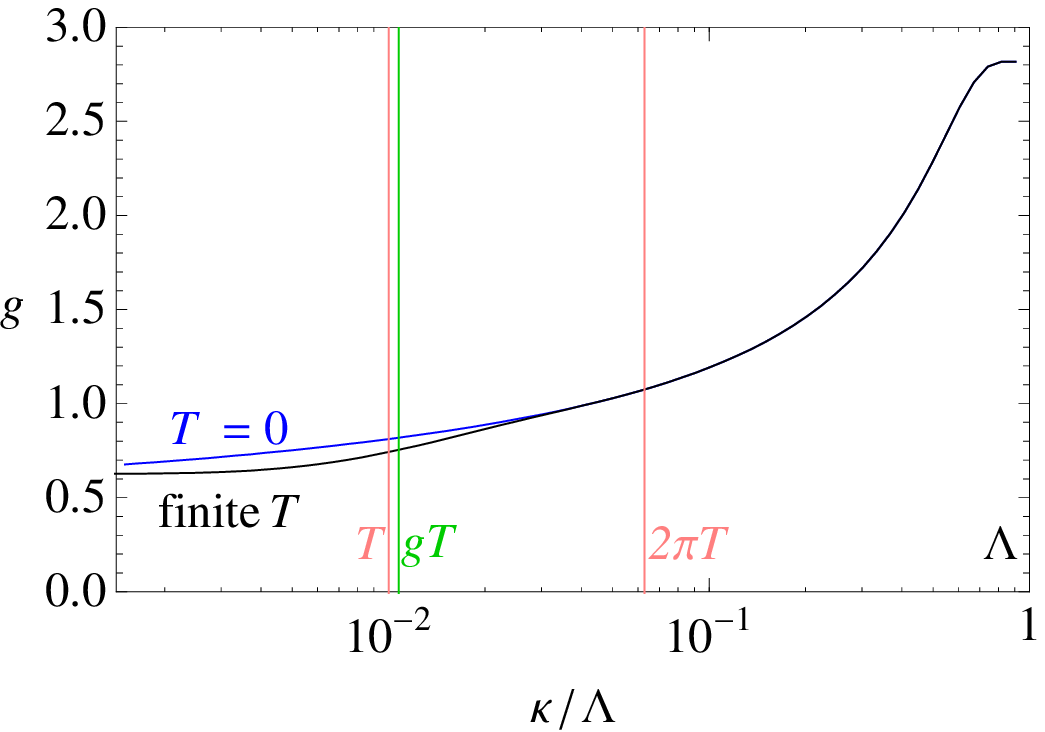}\hfill{}

\hfill{}\includegraphics[scale=0.65]{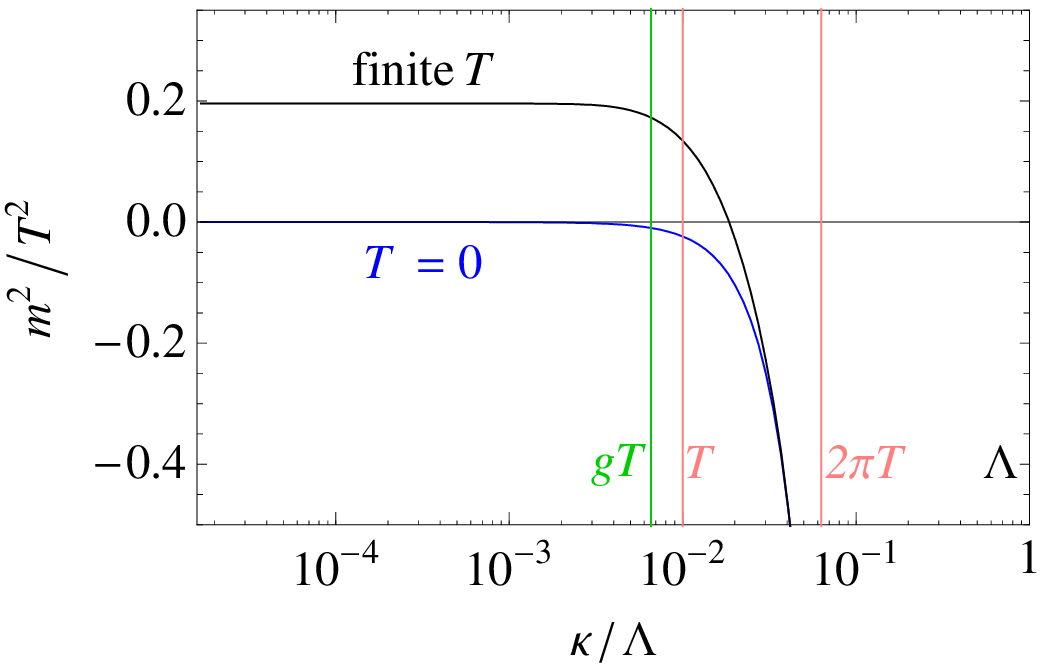}\hfill{}\includegraphics[scale=0.65]{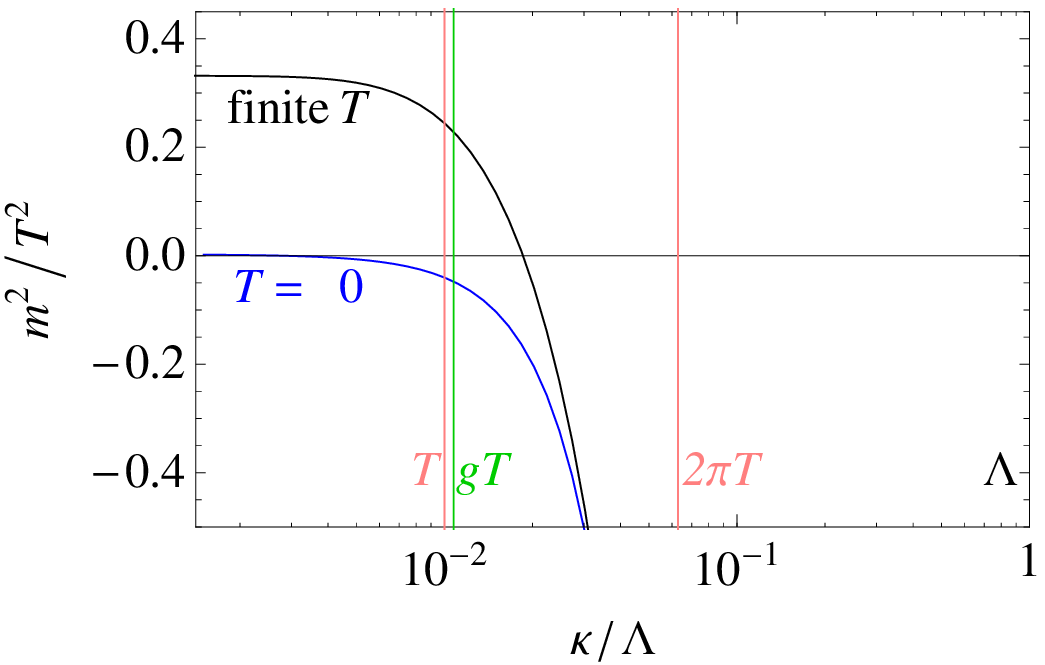}\hfill{}

\hfill{}\includegraphics[scale=0.65]{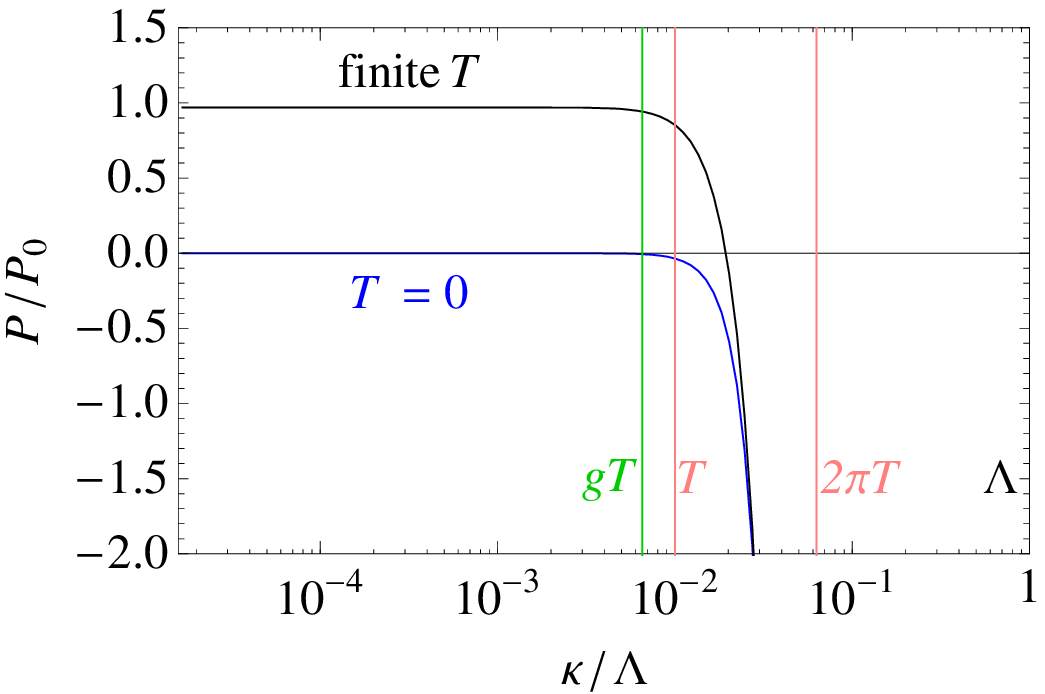}\hfill{}\includegraphics[scale=0.65]{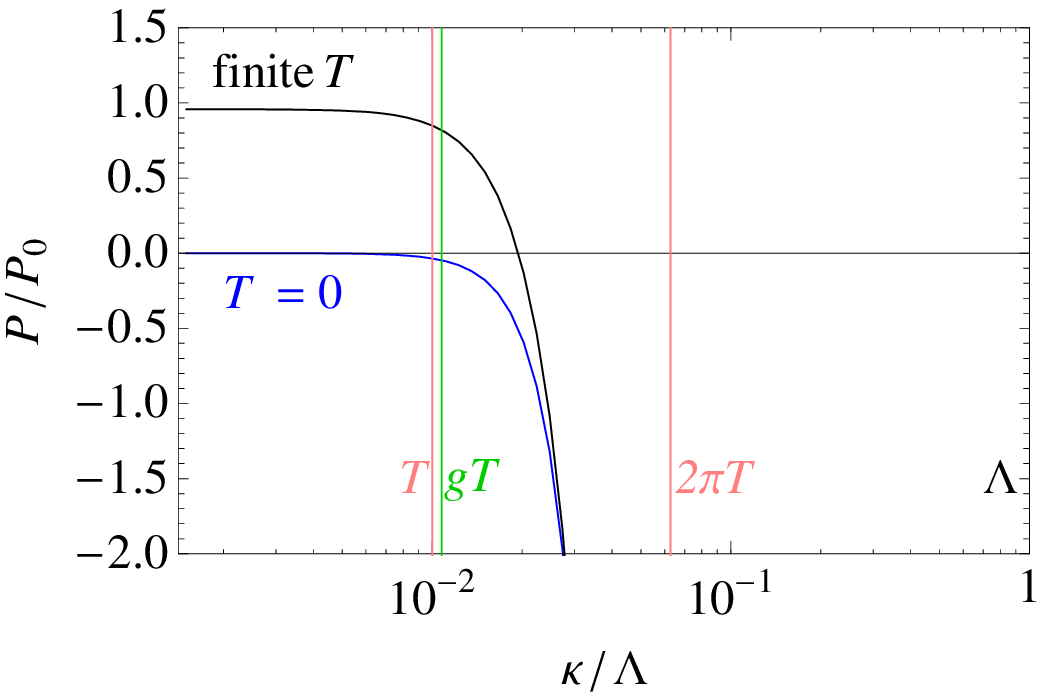}\hfill{}

\hfill{}\includegraphics[scale=0.65]{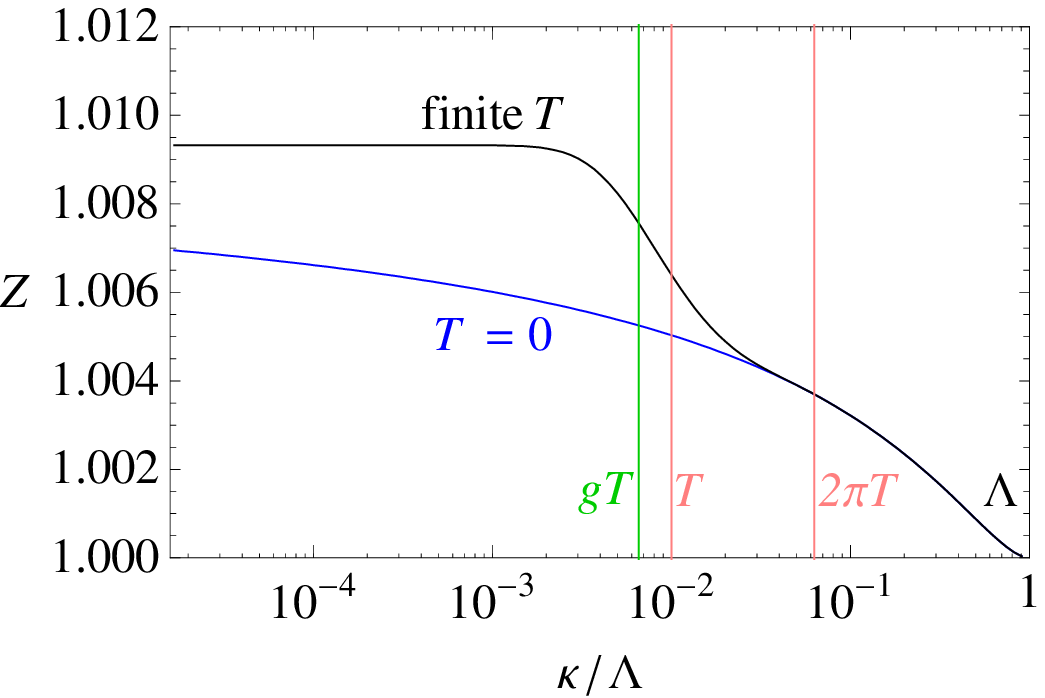}\hfill{}\includegraphics[scale=0.65]{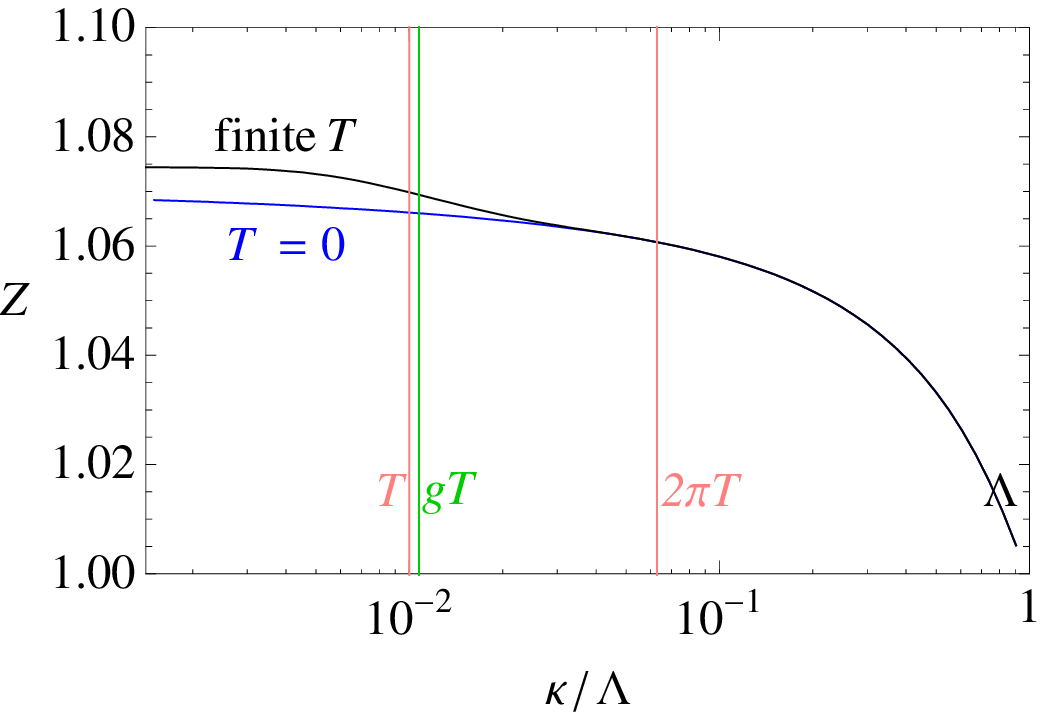}\hfill{}

\caption{({\it Color online}) Flow of various quantities in the BMW approximation as a function
of the flow parameter $\kappa$ at zero and finite temperature. The
left column shows results at small bare coupling ($g_\Lambda=0.88)$ while
the right column shows results at large bare coupling ($g_\Lambda=2.8)$.
Depicted are the flows of the quartic coupling, the mass, the pressure ($P_0=\pi^2 T^4/90$), and the Z-factor (top
to bottom). Vertical lines indicate the positions where $\kappa=gT$,
$T$, and $2\pi T$. \label{fig:flow}}

\end{figure}

We have adapted to finite temperature the numerical strategy 
that has been developed to solve the BMW equation at zero temperature in the context of $O(N)$ models \cite{Benitez:2009xg, Benitez:2010aa}.
This method puts propagator and potential on
a regular two-dimensional grid in $|\mathbf{q}|$ and $\rho$ variables and solves the
flow equations using the Euler method.
At finite temperature, the flow equation (\ref{eqforV}) involves
a summation over Matsubara frequencies. This summation is infeasible in practice 
at large values of $\kappa$, but at sufficiently large $\kappa\gg T$,
the corresponding vacuum integral constitutes the dominant contribution
and any thermal contribution is exponentially suppressed. Practically,
a switching temperature $T_{\mathrm{switch}}\gg T$ is introduced
such that, at $\kappa=T_{\mathrm{switch}}$, the numerical code switches from vacuum integration to thermal
integration with Matsubara frequencies. At this point, the values of the vacuum propagator $G_{\kappa}(q,\rho)$
which are stored on a two-dimensional grid $(|q|,\rho)$ have to be
interpolated in order to yield values on a three-dimensional grid $(\omega,|\mathbf{q}|,\rho)$
using cubic splines. Good results have been achieved on a $n_{q}\times n_{\rho}=90\times50$
grid with $\tilde{q}_{\mathrm{max}}=9$ and $\tilde{\rho}=5$ (in
dimensionless variables, see Eq.~(\ref{dimensionless})) with 30 Matsubara terms at the switching
temperature $T_{\mathrm{switch}}$ which is varied in the range $4\pi T\leq T_{\mathrm{switch}}\leq 6\pi T$. To speed up the calculation, the
number of Matsubara terms can be gradually reduced as the flow proceeds,
without affecting accuracy. Further details on the integration of the pressure flow are given in Appendix A.

\begin{figure}
\hfill{}\includegraphics[scale=0.75]{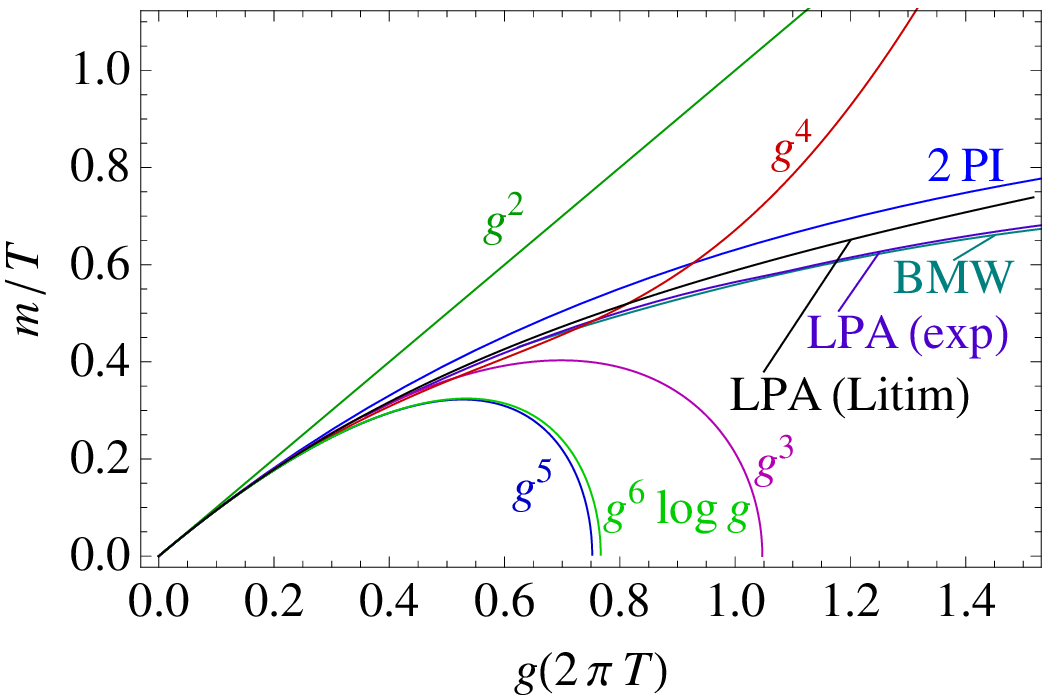}\hfill{}\includegraphics[scale=0.75]{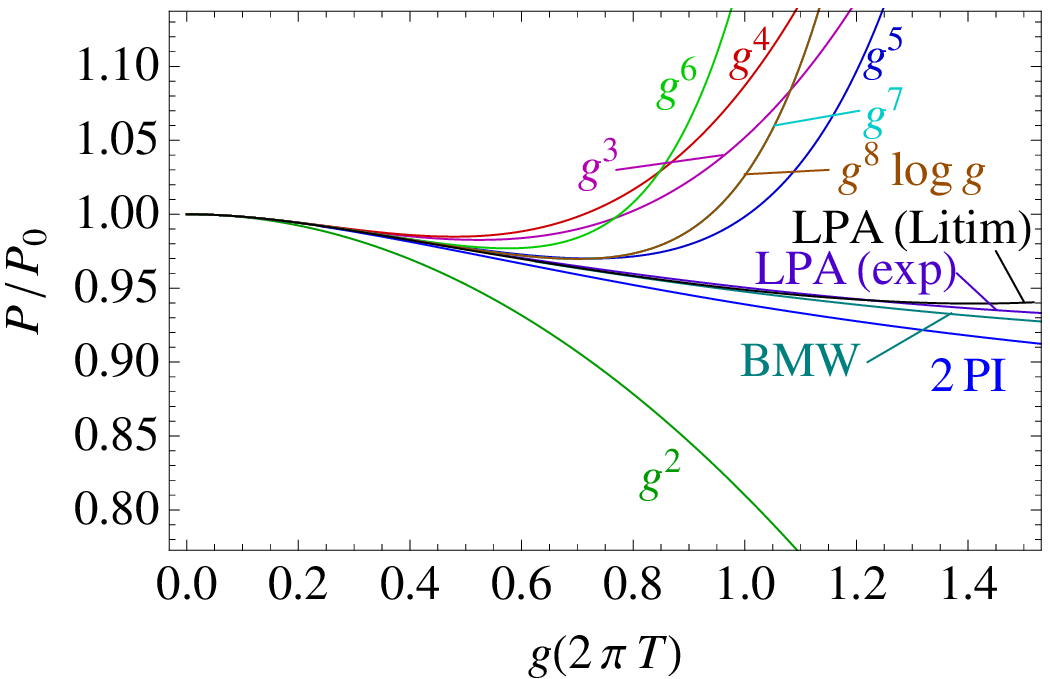}\hfill{}

\caption{({\it Color online}) Mass and pressure as function of the coupling. The various resummation and RG methods compared are 2 PI, LPA with exponential regulator (exp), LPA with Litim regulator, and BMW. Shown are also perturbative results through order $g^6 \log g$ for the mass and $g^8 \log g$ for the pressure. The $g^7$ and $g^8 \log g$ curves for the pressure almost lie on top of each other, as do the BMW and LPA (exp) curves for the mass.\label{fig:masspressure}}

\end{figure}

Figure \ref{fig:flow} shows the flow of various quantities (coupling constant, mass, pressure and $Z$-factor) in the
BMW approximation. (The pressure $P_0$ is that of the non-interacting scalar plasma, i.e., $P_0=\pi^2 T^4/90$.) These flows  are remarkably
similar to the corresponding ones obtained within the LPA approximation in Ref.~\cite{Blaizot:2006rj}.
At finite temperature, the flow starts to deviate from the vacuum
flow between $2\pi T$ and $T$, and stabilizes shortly below the scale $gT$ (with $g\equiv g(2\pi T))$. Note that $gT$ is the leading order value at small $g$ of the thermal mass of the excitations (see e.g. Fig.~\ref{fig:masspressure}): when $\kappa$ reaches this value the mass takes over the role of an infrared cut-off, which freezes the flow. 
For the bare value $g_\Lambda=0.88$ the flow of the coupling reaches a value
$g(\kappa=2\pi T)=0.66$, while for the larger bare coupling 
$g_\Lambda=2.8$ the value is $g(2\pi T)=1.1$.
Vertical lines indicate the positions of $\kappa = gT$, $T$, and $2\pi T$
for the temperature $T=10^{-2} \Lambda$.
The bottom figures displays the flow of 
the $Z$-factor, which is trivial  in the LPA approximation (where $Z=1$ for all values of $\kappa$).
It turns out that this factor deviates only moderately from 1 in the
BMW approximation. This indicates that the mixing of single particle excitations with multi-particle states is very mild, at least within the BMW approximation, and suggests that the quasiparticle picture is robust in this system. This could be confirmed by a complete calculation of the single particle spectral function, which is beyond the scope of this paper. 

Figure \ref{fig:masspressure} shows a comparison of different approximations to the mass and the pressure as a function
of the coupling $g(2\pi T)$. To obtain these plots, results similar to the ones shown in Fig.~\ref{fig:flow},
and  obtained for various bare couplings and ratios $T/\Lambda$ (where values up to $T/\Lambda=1/10$ have been used in order to extend the plot range to large values of $g(2 \pi T)$), have been combined into  single plots. 
Results obtained within the LPA and the BMW approximations to the NPRG  are compared to 
perturbation theory through order $g^6 \log g$ for the mass \cite{privateAndersen} and $g^8 \log g$  for the pressure \cite{Andersen:2009ct}, and also to the result of the 2-loop  2PI resummation from Ref.~\cite{Blaizot:2006rj}. The BMW results were all obtained with the exponential regulator (\ref{regulator1}), while the LPA results were obtained with this same  exponential regulator, as well as with the 3-dimensional Litim regulator used in Ref.~\cite{Blaizot:2006rj}. For a given choice of regulator, one sees in Fig.~\ref{fig:masspressure} that the difference between the LPA and the BMW results for both the mass and the pressure is tiny: one does not gain much, for the thermodynamics, in improving the treatment of long wavelength degrees of freedom and incorporating explicit momentum dependence in the self-energy. The stability of the BMW approximation scheme, comparable to that of the 2PI resummation,  should be contrasted with the wild oscillatory behavior of the successive orders of perturbation theory.

\begin{figure}
\hfill{}\includegraphics[scale=0.75]{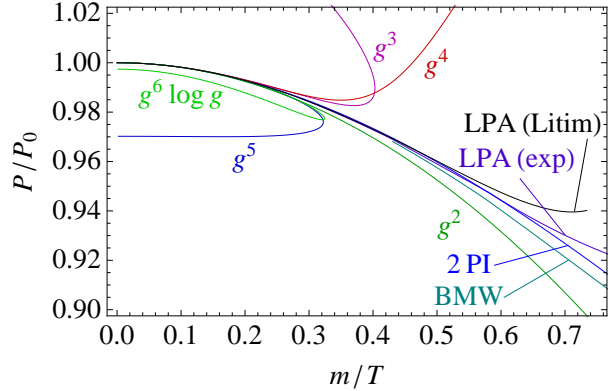}\hfill{}
\caption{({\it Color online}) Pressure as a function of the mass. The same curves are shown as in 3, but for the scale-independent function of pressure as a function of the thermal mass. The perturbative results are shown through order $g^6 \log g$.\label{fig:pressureovermass}}
\end{figure}

The plots in  Fig.~\ref{fig:masspressure} depend on the choice of the prescription
for obtaining the coupling at a particular scale (in this case at the scale $\mu=2\pi T$), and hence on the regulator: depending on the regulator, a given value of the coupling constant $g(2\pi T)$ will correspond to different bare coupling constants $g_\Lambda$, and hence to different initial conditions for the temperature flow. Together with the obvious dependence of the flow itself  on the regulator this will affect the final result.   The regulator dependence that is visible in the LPA results in  Fig.~\ref{fig:masspressure} is to be attributed in part to the way  the results are plotted, namely as a function of $g(2\pi T)$.  Most of such a ``scheme dependence'' can be eliminated by plotting only physical quantities. This is realized in 
Fig.~\ref{fig:pressureovermass} which shows the pressure as a function of the mass. In this combination, 
only physical quantities are compared,  that do no longer depend on a particular choice
of a scale at which one fixes the coupling $g$, as was the case in Fig.~\ref{fig:masspressure}.
While perturbation theory clearly breaks down above a certain ratio $m/T\gtrsim 0.3$,
RG methods and the 2PI approach give consistent results up to twice this value.
As noticed previously  \cite{Blaizot:2006rj}, the perturbative $g^2$ contribution
seems to be a surprisingly good approximation for the behavior at larger $m/T$
-- but only for the plotted ratio $(P/P_0)$ versus $(m/T)$: the $g^2$ curve representing $P/P_0$ in Fig.~\ref{fig:pressureovermass} is identical to the corresponding curve in Fig.~\ref{fig:masspressure} (right panel), since, as already mentioned, at leading order in the coupling constant, $m/T=g$. Thus the reason the curves corresponding to the various non-perturbative approximations are brought closer to the $g^2$ curve in Fig.~\ref{fig:pressureovermass} can be traced back to the fact that the mass increases much less rapidly than $g$ with increasing $g$, as clearly visible  in Fig.~\ref{fig:masspressure} (left panel).  Note that the dependence of the LPA results on the choice of the regulator remains present, although it is less important than in  Fig.~\ref{fig:masspressure}. We shall return to this issue shortly. 

\begin{figure}
\hfill{}\includegraphics[scale=0.75]{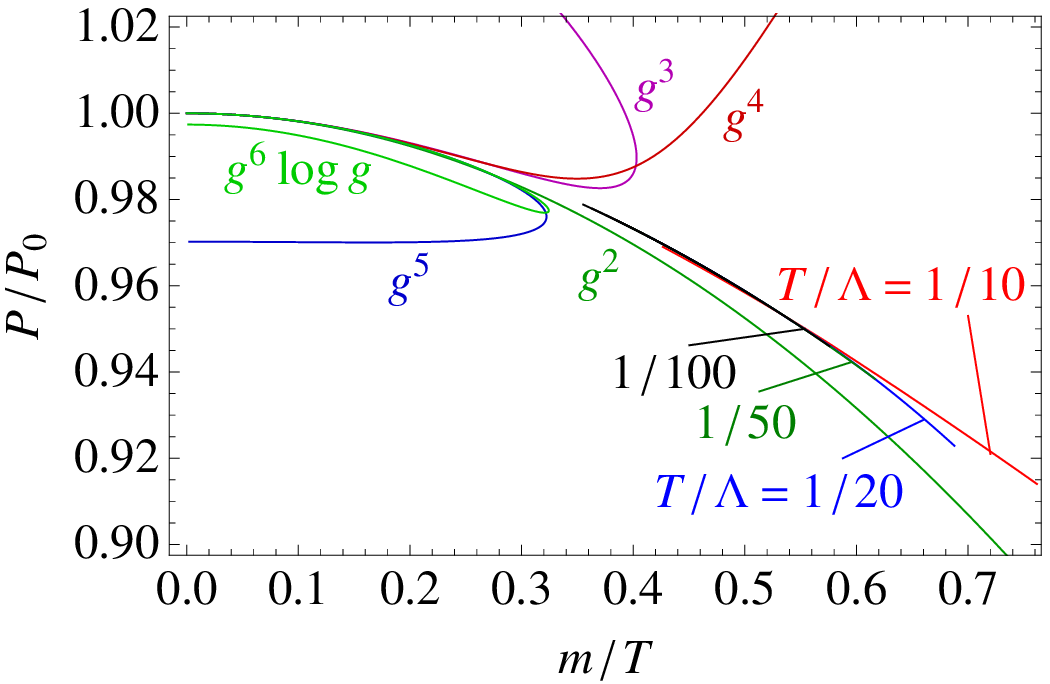}\hfill{}
\hfill{}\includegraphics[scale=0.75]{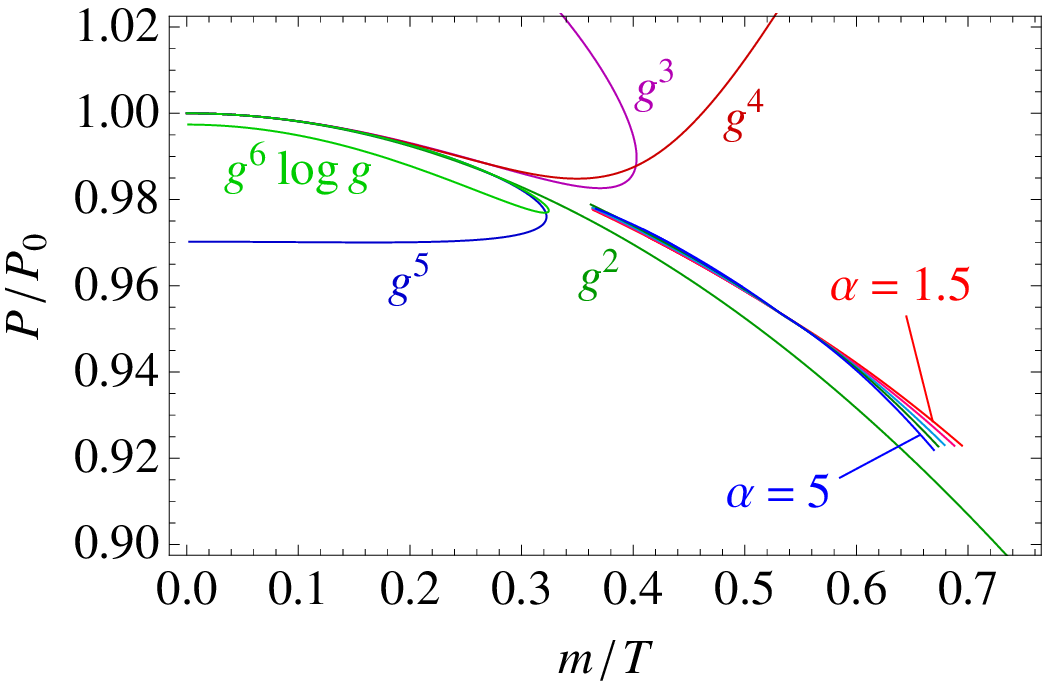}\hfill{}
\caption{ ({\it Color online}) Pressure as a function of the mass, within the LPA and the exponential regulator (\ref{regulator1}). Left: comparison of results for various
values of the temperature $T/\Lambda$. Right: comparison of results for various values of the exponential regulator parameter $\alpha=1.5, 2, 3, 4, 5$. In the right panel, the curves are extracted for $T/\Lambda = 1/20$ or smaller (so that the plots in this panel are  unaffected by a $T/\Lambda$ dependence).  \label{pressureTLambda}}
\end{figure}

We turn now to more technical aspects of the calculations, namely the dependence of the results displayed above on the choice of the temperature, or on the choice of the regulator. Consider first the dependence on the temperature, which is measured by its ratio $T/\Lambda$ to the microscopic scale $\Lambda$.  As $T/\Lambda$ increases and becomes close to 1, our numerical calculations loose accuracy for a variety of reasons, the main one being the following:  If $T$ is too close to $\Lambda$, $2\pi T$ may become bigger than $\Lambda$ and the whole procedure eventually collapses. One indeed assumes that the beginning of the flow is not affected by the temperature (so as to use the 4-dimensional integration procedures), and this assumes $\Lambda>2\pi T$ so that there is room for a 4-dimensional flow between $\Lambda$ and $2\pi T$, with $\kappa \approx 2\pi T$ the point where thermal fluctuations start to contribute to the flow.  This limitation is illustrated in Fig.~\ref{pressureTLambda}, for the case of the LPA (the phenomenon would be identical in the BMW approximation): the plot of the pressure versus the mass is independent of the temperature as long as $T/\Lambda\simle 1/20$, while the curve corresponding to $T/\Lambda= 1/10$, deviates slightly  from it. 

The right panel of Fig.~\ref{pressureTLambda} illustrates the dependence of the results on the value of the parameter $\alpha$ in the exponential regulator (\ref{regulator1}). This is achieved by repeating calculations for a set of values of $\alpha$ in the range $[1.5, 5]$. As one can see, the results are fairly insensitive to the value of $\alpha$, until the mass reaches a value of the order $m/T\simeq 0.5\div 0.6$, where a sizable dependence starts to be visible. Note that this is the value of the mass where we observed possible cut-off effects when the temperature is too large. In the present case, the temperature is not to blame. However cut-off effects may show up in the fact that integrals are done with $\Lambda$ as ultraviolet cut-off. In most cases this is redundant and does not interfere with the regulator (whose derivative with respect to $\kappa$ also provides an ultraviolet cut-off). However in the mass region $m/T\approx 0.5\div 0.6$, the two cut-offs interfere, and produce a sensitivity of the results to the regulator: indeed a larger $\alpha$ allows larger momenta. 

\begin{figure}
\hfill{}\includegraphics[scale=0.75]{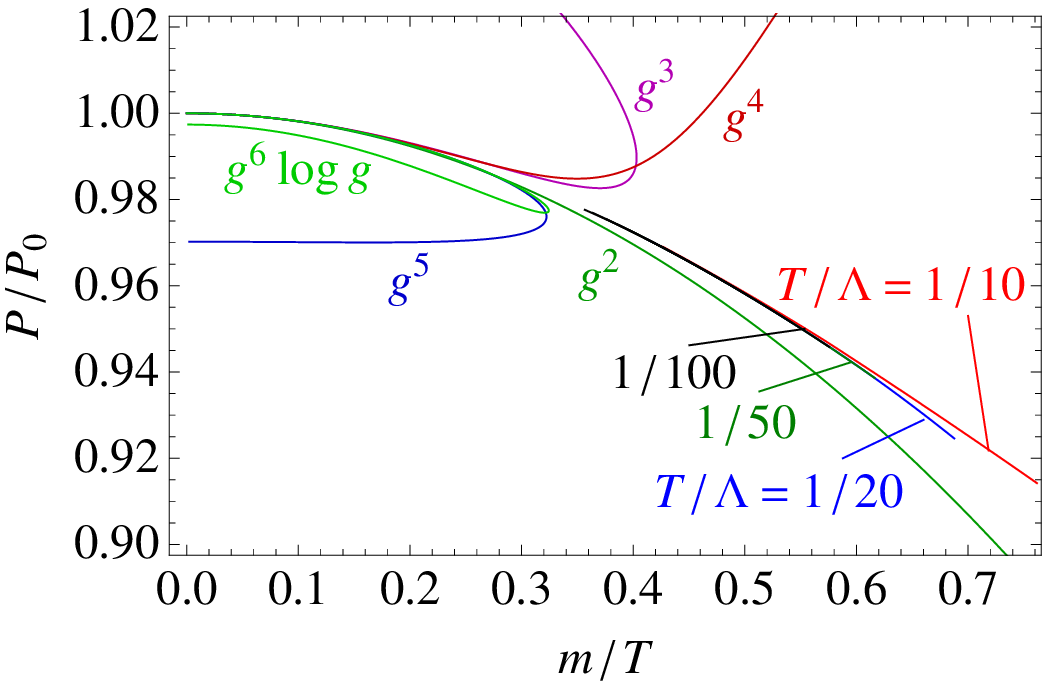}\hfill{}
\hfill{}\includegraphics[scale=0.75]{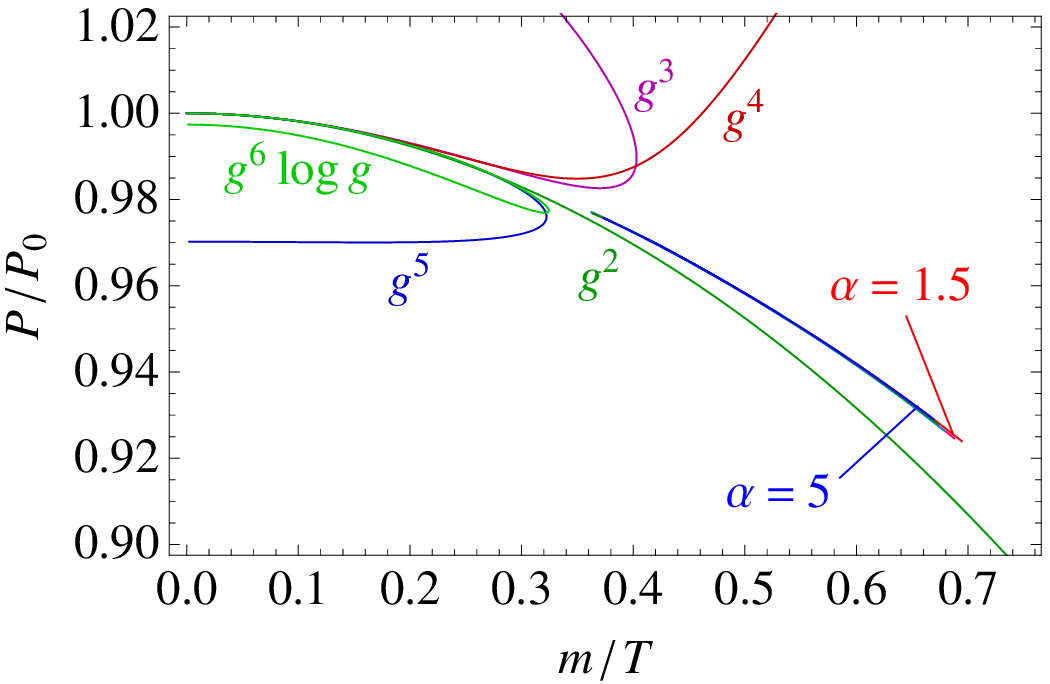}\hfill{}
\caption{({\it Color online}) Pressure as a function of the mass. Comparison of LPA plots at various
values of the regulator parameter $\alpha=1.5,3,4,5$.  \label{pressureTLambda2}}
\end{figure}

\begin{figure}
\hfill{}\includegraphics[scale=0.75]{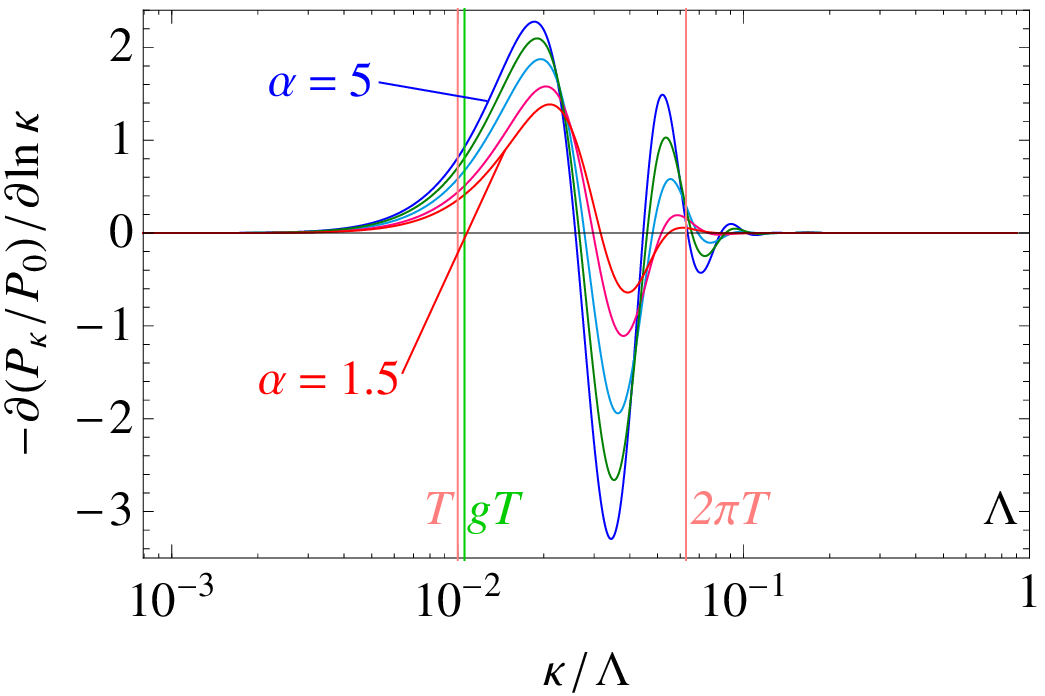}\hfill{}\includegraphics[scale=0.75]{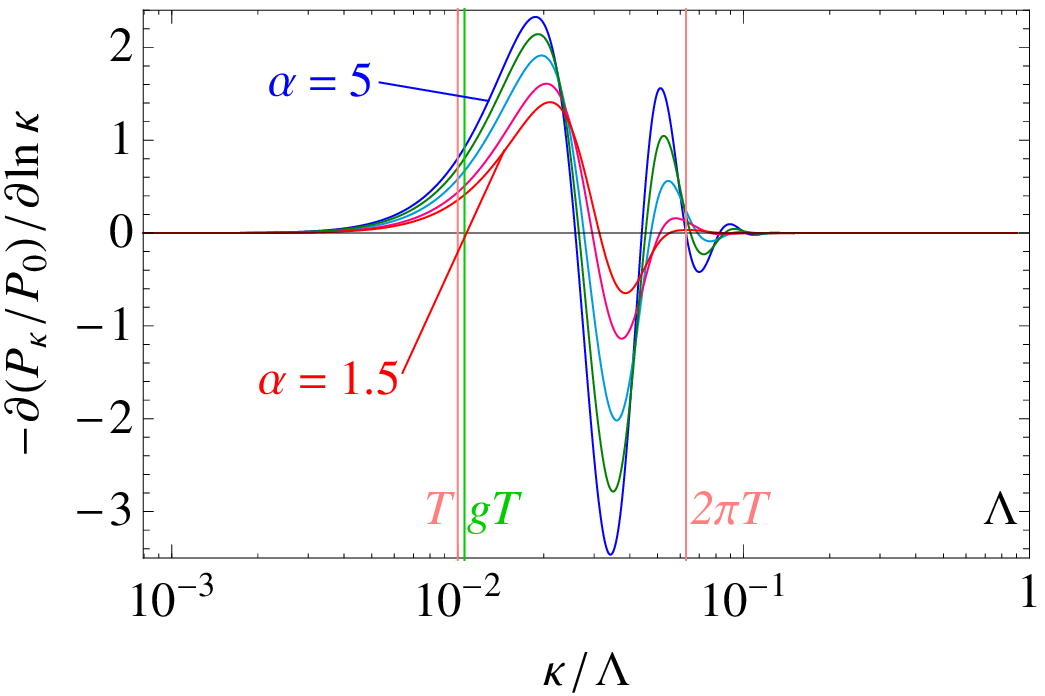}\hfill{}

\caption{({\it Color online}) Derivative of the flow of the potential difference between thermal and vacuum contribution.
Comparison of LPA contributions for the exponential regulator (\ref{regulator1}) (left) or (\ref{regulator2}) (right) for various values of the
parameter $\alpha=1.5,2,3,4,5$. The larger the value of $\alpha$,
the more oscillations are introduced through the regulator. \label{fig:potentialflow}}

\end{figure}

To investigate this further, we have considered the other regulator mentioned in Sect.~\ref{sec:bmw}, namely, Eq.~(\ref{regulator2}). This regulator has the property to cut-off more efficiently the large momenta (because of the presence of the $q^4$ term). And indeed this is what one sees in Fig.~\ref{pressureTLambda2}. The left panel indicates that a higher temperature can be reached without affecting the results. The right panel shows that the dependence on the regulator parameter $\alpha$ has basically disappeared. A further confirmation comes form the study of the dependence of the results on the temperature $T_{\mathrm{switch}}$ at which one changes from the 4-dimensional to the 3-dimensional flow: essentially no such dependence  is observed with the second regulator as long as $T_{\mathrm{switch}}\simge 4\pi$. One should emphasize however that the difference between the two regulators is   tiny. Thus for instance, the plots in Fig.~\ref{fig:potentialflow} show the derivative of the pressure as a function of $\kappa$. The oscillatory behavior at the beginning of the flow is generic for regulators that cut-off the sum over the Matsubara frequencies, and the particular oscillations exhibited in Fig.~\ref{fig:potentialflow}  may be understood from the analytical structure of the regulator analyzed in detail in Appendix B. One sees clearly that the larger $\alpha$ the larger are the oscillations. On the other hand, there is hardly any visible difference between the two panels corresponding to the two different regulators. Only if one zooms in a lot, can one see that (\ref{regulator2}) is slightly better in cutting off $\partial P_\kappa$ at larger values of $\kappa$  than (\ref{regulator1}).
The plots show curves for $g_\Lambda=2.8$, corresponding to $g(2\pi T)=1.1$ and $T=10^{-2}\Lambda$ (that is, the same parameters as the right column of Fig.~\ref{fig:flow}).

\section{Conclusions}

In this paper we have applied a powerful  approximation scheme to solve the NPRG equations and we have calculated various thermodynamical quantities for a
scalar field theory at finite temperature, in a range of values of the coupling constant 
covering both weak and strong coupling regimes.
  Of course, 
because of the presence of the Landau pole, arbitrarily large values 
of the coupling cannot be reached. However the range of values of $g$ 
that can be explored allowed us to demonstrate the stability of the results as successive orders in the approximation scheme are taken into account, in sharp contrast to the strict expansion in terms of the coupling constant. A similar stability also emerges in other non-perturbative schemes, such as ``screened perturbation theory"  \cite{Andersen:2008bz}, or the 2PI effective action \cite{Berges:2004hn} (see also \cite{Blaizot:2003iq}).

In fact, the results that we have obtained using the BMW approximation  differ very little from those that we obtained previously using the LPA \cite{Blaizot:2006rj}. This may be attributed to the fact that the thermal mass provides an infrared cut-off that reduces the contributions to the pressure of the long wavelength modes, which are the most strongly coupled. The thermal mass  also provides a threshold that hinders the mixing with complex multiparticle configurations, making the quasiparticles well defined. This latter aspect is explicitly verified by the small deviation of the field normalization from unity, as obtained within the BMW approximation. These two effects conspire to give the approximation scheme a remarkable stability, and contribute to the robustness of the quasiparticle picture. 

We 
have also compared the results obtained within the BMW scheme with those of the 
simple 2-loop 2PI resummation used in \cite{Blaizot:2006rj}: both methods lead to very similar results in 
the extrapolation to strong coupling. This is not too surprising, given the agreement already observed between the LPA and the 2PI method  in Ref.~\cite{Blaizot:2006rj}. The stability of the 2PI scheme itself can be assessed from  the 3-loop calculation of Ref.~\cite{Berges:2004hn}. The detailed comparison between the latter calculation and ours is not straightforward however because of the different renormalization schemes used in the two cases.  However, the main message is essentially the same: the main qualitative difference between the 2-loop and 3-loop calculations is the presence of momentum-dependent self-energies in the latter, in contrast to simple mass terms in the former. The small difference observed between the results in the two cases, corroborates the  conclusions that follow from our NPRG analysis about the robustness of the quasiparticle picture. In fact the stability of the results lead us to conjecture that any corrections to them are presumably very small. It would be interesting to have lattice calculations allowing us to test this conjecture for values of the couplings where perturbation theory breaks down. 

\acknowledgments
The authors gratefully acknowledge the hospitality of the ECT* in Trento, where part of the work reported here was carried out, at different periods of time. Special thanks are due to Ramon M\'endez-Galain for discussions at an early stage of this project. NW acknowledges support from the program PEDECIBA (Uruguay), and JPB  support form the Austrian-French  program Amadeus 19448YA. We also thank Jens O. Andersen for useful correspondence concerning his high order perturbative calculations.

\appendix

\section{Numerical integration of the pressure flow}

\begin{figure}
\hfill{}\includegraphics[scale=0.5]{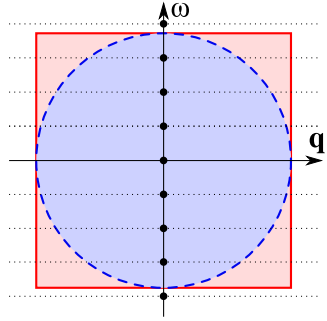}\hfill{}

\caption{{\it(Color online)} Integration domains for thermal and vacuum pieces.
For the zero-temperature piece, Euclidean invariance imposes a four-dimensional
sphere for a given cutoff $\Lambda$ (blue dashed circle), while the
summation over Matsubara frequencies up to a maximum frequency independent of the three momentum,  implies a four-dimensional cylinder
(red solid square). For finite $\Lambda$, this mismatch can give
rise to spurious contributions to the pressure. \label{fig:integrationdomains}}

\end{figure}

The calculation of the pressure involves  a delicate cancellation between
thermal and vacuum contributions at large momenta. The thermal pressure is obtained from the minimum of the effective potential, after subtracting the vacuum contribution, as indicated in Eq.~(\ref{pressuredef}). It is of the form $P=\int_{0}^{\Lambda}d\kappa\left(\partial_{\kappa}V_{\kappa}^{T}-\partial_{\kappa}V_{\kappa}^{T=0}\right)$, where
the vacuum ($T=0$) integral with vacuum propagator is subtracted from the
thermal sum with thermal propagator: \begin{equation}
\partial_{t}V_{\kappa}^{T}-\partial_{t}V_{\kappa}^{T=0}=\frac{1}{2}T\sum_{n=-n_\mathrm{max}}^{n_\mathrm{max}}\int^\Lambda\frac{d^{3}\mathbf{q}}{(2\pi)^{3}}\partial_{t}R_{\kappa}(q)G_{\kappa}^{T}(q,\rho)-\frac{1}{2}\int^\Lambda\frac{d^{4}q}{(2\pi)^{4}}\partial_{t}R_{\kappa}(q)G_{\kappa}^{T=0}(q,\rho)\label{eq:potentialdiff1}\end{equation}
with $\int^\Lambda d^{3}\mathbf{q}\equiv \int_{\mathbf{q}^2\leq \Lambda^2} d^{3}\mathbf{q}$
and  $\int^\Lambda d^{4}q\equiv \int_{q_0^2+\mathbf{q}^2\leq \Lambda^2} d^{4}q$.
As can be seen in Fig.~\ref{fig:integrationdomains}, the two integration
domains do not match. For large cutoffs $\Lambda$, this mismatch
may give rise to spurious contributions to the pressure. 

One solution could seem to be to restrict the range of the Matsubara
sum to a 4-dimensional sphere. This turns out to be problematic for
the following reason: Good convergence properties are only obtained
if sufficiently many Matsubara terms are summed up, but at large values
of $\mathbf{q}$, only few Matsubara terms would contribute if 
the restriction to a 4-dimensional sphere were imposed.

It is better then to fix the mismatch in an alternative way, i.e., by extending the  vacuum
integration domain so that it matches the domain of the Matsubara
summation. That is, we write \begin{eqnarray}
\partial_{t}V_{\kappa}^{T}-\partial_{t}V_{\kappa}^{T=0}&=&\frac{1}{2}T\sum_{n=-n_\mathrm{max}}^{n_\mathrm{max}}\int^\Lambda\frac{d^{3}\mathbf{q}}{(2\pi)^{3}}\partial_{t}R_{\kappa}(q)G_{\kappa}^{T}(q,\rho)
\nonumber\\ &-&\frac{1}{2}\int_{-q_{0,\mathrm{max}}}^{q_{0,\mathrm{max}}}\frac{dq_{0}}{2\pi}\int\frac{d^{3}\mathbf{q}}{(2\pi)^{3}}\partial_{t}R_{\kappa}(q)G_{\kappa}^{T=0}(q,\rho),\label{eq:potentialdiff2}\end{eqnarray}
for $q_{0,\mathrm{max}}=2\pi n_\mathrm{max} T$.
Practically, only the pressure is sensitive to this correction. At
each integration step in $t$ direction, the vacuum integral is calculated
twice: Once as in Eq.~(\ref{eq:potentialdiff1}) to follow the vacuum
flow, and additionally as in Eq.~(\ref{eq:potentialdiff2}) to obtain
the contribution for the pressure. The values of $G_{\kappa}^{T=0}(q,\rho)$
that are only known on a grid in $(|q|,\rho)$ coordinates have to
be interpolated to obtain values on a three-dimensional grid $(q_{0},|\mathbf{q}|,\rho)$.
Since for each temperature the switching point between 4D vacuum integration
and 3D Matsubara summation varies, the vacuum piece for Eq.~(\ref{eq:potentialdiff2})
has to be calculated for each temperature separately.

\section{Oscillatory behavior of the flow}

As we have seen in Sect.~\ref{sec:numerical}, the flow of the pressure exhibits an oscillatory behavior at the beginning of the flow. This generically occurs whenever the sum over the Matsubara frequencies is limited to a finite number of terms.  We show in this Appendix that the oscillations seen in Fig.~\ref{fig:potentialflow} at the beginning of the flow can be understood from  the analytic structure of the particular regulator that we are using, and can be simply calculated for small values of $\kappa$. The analysis is performed within the LPA with the exponential regulator (\ref{regulator1}) with $\alpha=1$. (Note that much wilder oscillations than the one discussed here are observed when one uses a ``hard'' regulator, such as the Litim regulator \cite{Litim:2001up}.)

Instead of performing the sum over Matsubara frequencies in Eq.~(\ref{eqforV}) explicitly,
we can convert it into a contour integration in the following way:\begin{eqnarray}\label{B1}
T\sum_{n=-\infty}^{\infty}f(q_{0}=i\omega_{n}) & = & \frac{1}{2\pi i}\int_{-i\infty}^{i\infty}dq_{0}\frac{1}{2}\left[f(q_{0})+f(-q_{0})\right]\label{eq:sum}\\
 &  & +\frac{1}{2\pi i}\int_{-i\infty+\epsilon}^{i\infty+\epsilon}dq_{0}\left[f(q_{0})+f(-q_{0})\right]\frac{1}{e^{ q_{0}/T}-1}.\nonumber \end{eqnarray}
The resulting integrals can then be calculated by  closing the vertical contours with semi circles
of infinite radii, and summing over the encircled residues.

\begin{figure}
\begin{centering}
\hfill{}\includegraphics[scale=0.55]{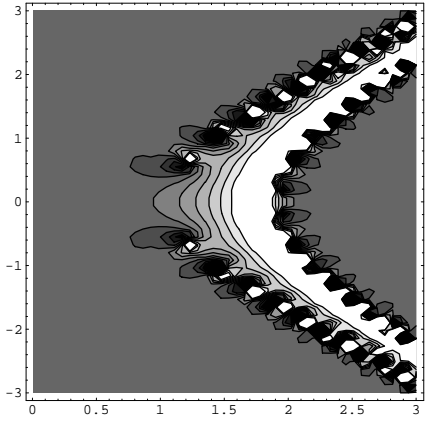}\hfill{}
\par\end{centering}

\caption{ Contour plot in the complex $q_{0}$ plane of the integrand of Eq.~(\ref{eqforV}),
for $|\mathbf{q}|=1$, $m=0.01$, and $\kappa=0.5$. The left bow
of poles corresponds to trivial poles whose residues can be calculated
analytically. The right bow of poles can only be obtained numerically. \label{contourplot}}

\end{figure}

It is convenient to group the poles of the integrand in Eq.~(\ref{eqforV})   into two categories (see Fig.~\ref{contourplot}). There are the poles associated with the regulator, which we shall call trivial poles, and those of the propagator, that is the zeros, of \begin{equation}
G_{\kappa}^{-1}(q,\rho)=q^{2}+m_{\kappa}^{2}+R_{\kappa}(q).\label{eq:propagator}\end{equation}

The positions of the trivial poles are  determined by  \begin{equation}
\exp\left[\left(-q_{0}^{2}+\mathbf{q}^{2}\right)/\kappa^{2}\right]-1=0.\end{equation}
This is satisfied for $
\left(-q_{0}^{2}+\mathbf{q}^{2}\right)/\kappa^{2}=2\pi in $,  with $n\in\mathbb{N}$, with the solution $
q_{0}=\pm\sqrt{\mathbf{q}^{2}-2\pi in\kappa^2�}$.
For $|\mathbf{q}|=0$, these poles lie at (for $\mbox{Re}q_{0}>0$) $ 
q_{0}=\kappa\left(1\pm i\right)\sqrt{\pi n}.$ 
For $|\mathbf{q}|\neq0$, the poles lie along a hyperbola. For completeness, we give 
the explicit real and imaginary parts of $q_0$: \begin{equation}
q_{0}=\sqrt{\sqrt{\left(\pi n\right)^{2}\kappa^{4}+\frac{\mathbf{q}^{4}}{4}}+\frac{\mathbf{q}^{2}}{2}}\pm i\sqrt{\sqrt{\left(\pi n\right)^{2}\kappa^{4}+\frac{\mathbf{q}^{4}}{4}}-\frac{\mathbf{q}^{2}}{2}}.\label{eq:polepositions}\end{equation}

The non-trivial poles can only be found by numerically by solving\begin{equation}
-q_{0}^{2}+\mathbf{q}^{2}+m_{\kappa}^{2}+R_{\kappa}(-q_{0}^{2}+\mathbf{q}^{2})=0.\end{equation}
It turns out that the non-trivial poles lie to the right of the trivial
poles in the complex plane. Due to the statistical factor in Eq.~(\ref{B1}), the corresponding
residues are therefore exponentially suppressed. In fact, 
the first few trivial poles are sufficient to accurately approximate Eq.~(\ref{eqforV}) for large values of $\kappa$.

Let us then calculate the residue for the integrand of Eq.~(\ref{eqforV})
at the trivial pole positions (\ref{eq:polepositions}). First we
note that we can write the derivative of Eq.~(\ref{regulator}) for the regulator (\ref{regulator1}) as \begin{equation}
\partial_{t}R_{\kappa}(q)=\frac{2}{\kappa^{2}}\left(q^{2}R_{\kappa}(q)+R_{\kappa}^{2}(q)\right).\end{equation}
Since $R_{\kappa}(q)$ diverges at the trivial pole, the propagator
(\ref{eq:propagator}) is dominated by the regulator, and the
residue is independent of the mass $m_{\kappa}$.  We have
\begin{eqnarray}
\underset{q_{0}\rightarrow q_{0}(n)}{\mbox{Res}}\frac{1}{e^{q_{0}/T}-1}\mathbf{q}^{2}\partial_{t}R_{\kappa}(q)G_{\kappa}(q,\rho) & = & \underset{q_{0}\rightarrow q_{0}(n)}{\mbox{Res}}\frac{1}{e^{q_{0}/T}-1}\frac{2\mathbf{q}^{2}}{\kappa^{2}}R_{\kappa}(q)\nonumber \\
 & = & -\frac{1}{e^{q_{0}(n)/T}-1}\mathbf{q}^{2}\frac{-q_{0}^{2}(n)+\mathbf{q}^{2}}{q_{0}(n)},\end{eqnarray}
with the poles given by $
q_{0}(n)\equiv\pm\sqrt{\mathbf{q}^{2}-2\pi in\kappa^2�}$.  The result follows easily from\begin{equation}
\underset{q_{0}\rightarrow q_{0}(n)}{\mbox{Res}}R_{\kappa}(q)=-\kappa^{2}\frac{-q_{0}^{2}(n)+\mathbf{q}^{2}}{2q_{0}(n)}.\end{equation}
The oscillatory behavior follows from the distribution function $1/(e^{q_{0}/T}-1)$
with complex number $q_{0}$.

\begin{figure}
\begin{centering}
\hfill{}\includegraphics[scale=0.75]{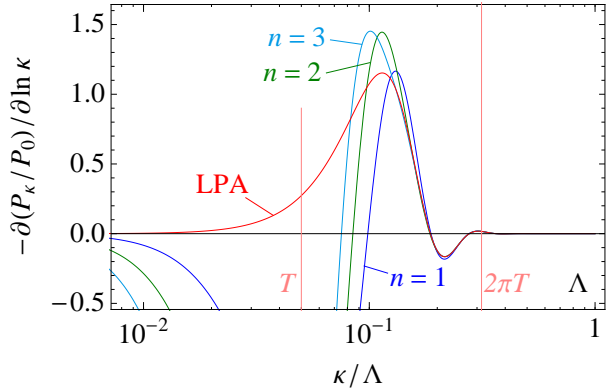}\hfill{}
\par\end{centering}

\caption{{\it(Color online)} Numerical solution (LPA) of the thermal contribution of Eq.~(\ref{eqforV}),
compared to the first three terms of (\ref{eq:analytic}) with $n=1$, 2, or 3,
as a function of $t=\ln\kappa$,
for $T=\Lambda/20$ and $m=\Lambda/1000$. The numerical solution is calculated through
the difference between an explicit summation over Matsubara frequencies
and a 4D integration. For larger values of $\kappa$, the pole residues
nicely reproduce the oscillating behavior. For smaller values of $\kappa$,
the analytic solution does not converge to the numerical result, because
the non-trivial numerical poles are missing in the calculation. \label{oscillations}}

\end{figure}

The thermal contribution is directly given by the second line of (\ref{eq:sum}).
One can write the contribution of the trivial poles as (instead of
adding the residues above and below the real axis, we simply can take
two times the real value of one of the residues; another factor 2
comes from $f(q_{0})+f(-q_{0})$; the factor $4\pi=\int d\Omega$)\begin{equation}
\partial_{t}\left[V_{\kappa}^{T}(\rho)-V_{\kappa}^{T=0}(\rho)\right]\approx2\frac{1}{2}\frac{4\pi}{(2\pi)^{3}}\sum_{n\geq1}\int_{0}^{\infty}dq\,2\mbox{Re}\left[\frac{1}{e^{q_{0}(n)/T}-1}q^{2}\frac{-q_{0}^{2}(n)+q^{2}}{q_{0}(n)}\right]\end{equation}
or explicitly:\begin{equation}
\partial_{t}\left[V_{\kappa}^{T}(\rho)-V_{\kappa}^{T=0}(\rho)\right]\approx\frac{1}{2\pi^{2}}\sum_{n\geq1}\int_{0}^{\infty}dq\,2\mbox{Re}\left[\frac{1}{e^{\sqrt{q^{2}-2\pi i\kappa^2�}/T}-1}q^{2}\frac{2\pi in\kappa^2�}{\sqrt{q^{2}-2\pi in\kappa^2�}}\right]\label{eq:analytic}\end{equation}
For large $\kappa$ this gives a good approximation to the thermal
pressure contribution, as can be seen in Fig.~\ref{oscillations}. Taking only the first term $n=1$ already gives
a good approximation for large values of $\kappa$. %


\end{document}